# Quantum Divergence and Topological Edge Diagnostics via Levitov Full Counting Statistics

Maolin Bo*,Xiang Chen, Siyu Liu, Han Lu, Yunhu Zhu, Zhongkai Huang, Chuang Yao

*Key Laboratory of Extraordinary Bond Engineering and Advanced Materials Technology (EBEAM) of Chongqing, School of Materials Science and Engineering, Yangtze Normal University, Chongqing 408100, China*

*Corresponding author: Maolin Bo (E-mail address: bmlwd@yznu.edu.cn)

## Abstract

We propose a differential full-counting-statistics protocol for mesoscopic transport. Additionally, we compare terminal Fano factors and noise cumulants between gate configurations at matched $|k_1|$, instead of inferring a bulk divergence sensor from a single absolute F. The operational excess $\Delta\chi_{\text{hidden}} = \chi_\Gamma - \chi_{\text{ref}}$ is a bookkeeping definition on the Lesovik backbone; it is illustrated analytically for a two-channel factorization via a zero-temperature geometry scan. Secondary benchmarks show that a two-dimensional lattice non-equilibrium Green's function calculation yields sub-Poissonian Fano factors, whereas Kumar's low-temperature quantum-point-contact calibration validates the numerical implementation.

# 1. Introduction

The statistical description of charge transport and current fluctuations in quantum coherent systems remain central problems in condensed matter and mesoscopic physics. Full counting statistics (FCS)—pioneered by Levitov et al.—provides an effective cumulant-generating-function language for charge transfer through open conductors[1-3], whereas integer quantum Hall and Chern insulator edges exhibit quantized conductance and sub-Poissonian shot noise[4-6]. Despite extensive research, no measurement-domain language clearly distinguishes between closed-boundary, bulk (Gaussian), and open-terminal counting objects within the same Lesovik backbone. Cryogenic quantum-point-contact and integer quantum Hall effect (IQHE) experiments[7-9] firmly established the phenomenological landscape; however, none directly compared the bulk quantity $\chi_{\mathrm{div}}$ with the terminal quantity $\chi_{\Gamma}$. To fill this gap, this study introduces a measurement-domain taxonomy that distinguishes among direct measurements, model-dependent references, and consistency verifications.

Experimentally, current noise spectroscopy provides quantitative benchmarks for observables targeted by this framework. Shot-noise measurements in quantum point contacts and mesoscopic conductors confirmed the Levitov–Lesovik reduction of the Fano factor below the Poisson limit of $F = 1$[3-4, 7]. In topological systems, shot-noise measurements on disordered helical-edge transport in inverted-band HgTe-based quantum wells reported sub-Poissonian Fano factors of $0.1 < F < 0.3$.[8] Graphene quantum Hall–superconductor junctions reported low-bias Fano factors of $F_{\mathrm{S}} < 0.5$ along with quantized conductance[10-11]. Integer quantum Hall edges established Chern-number-dependent conductance plateaus of $G = Ce^2/h$,[5] and edge-channel spectroscopy at cryogenic temperatures confirmed chiral fermion energy distributions in integer quantum Hall-effect beam splitters[6]. However, companion interacting-edge noise thermometry remains theoretical[12]. At room temperature, atomic-scale Au junctions exhibit shot-noise suppression near conductance quanta[9], while electrochemical electrode–electrolyte interfaces present Faradaic shot noise, which establishes a fundamental quantification limit[13-15]. These experiments establish clear qualitative comparison windows: sub-Poisson ($F < 1$)[3-4, 7-8], HgTe edge transport[8, 16-17], and room-temperature junction suppression[9]. Therefore, quantized edge conductance, Chern-dependent conformal field theory reference spectra, and preset-dependent Fano factors are benchmark values under the stated parameters. Meanwhile, open-terminal vs. bulk-divergence counting ($\chi_\Gamma$ vs. $\chi_{\text{div}}$) remains a proposed multiterminal diagnostic.

In this study, we construct a framework beginning with the Dirac field and global U(1) symmetry. Noether's theorem yields the conserved current operator $\hat{J}$, satisfying

$$\frac{\partial \hat{\rho}_e}{\partial t} + \nabla^T \cdot \hat{J} = 0 . \tag{1}$$

Promoting the classical Gauss divergence theorem to the operator level and introducing a spatial counting field $\lambda(r)$, we obtain an exact identity between volume and surface flux operators. When $\lambda$ is uniform, the bulk-divergence-generating functional $\chi_{\text{div}}$ coincide with the closed-boundary flux functional $\chi_{\partial V}$ . This equality is referred to as the quantum divergence theorem, which is stated for a spatially uniform counting field on a closed boundary. Moreover, it is not a model-independent nonperturbative quantum field theory for generic chiral edge systems with anomaly corrections beyond current conservation.

The verifiable core of this work is a differential FCS protocol that involves comparing terminal Fano factors and cumulants across gate configurations at matched $|\boldsymbol{\kappa}_1|$, instead of inferring bulk divergence from a single $F$. The open-terminal excess $\Delta\chi_{\text{hidden}} = \chi_\Gamma - \chi_{\text{ref}}$ is defined as a bookkeeping quantity on the Lesovik backbone, with $\chi_{\text{ref}}$ representing a closed-surface reference on the same mode set. The closed-boundary identity $\chi_{\partial V} = \chi_{\text{div}}$ for a uniform $\lambda$ is a consistency verification under current conservation, not a dynamical theorem; the transport content lies in the deviation of $\chi_\Gamma$ from this reference, as summarized in **Table 1**. Secondary benchmarks show that a two-dimensional (2D) lattice non-equilibrium Green's function calculation yields sub-Poissonian Fano factors, whereas Kumar's low-temperature quantum-point-contact calibration validates the numerical implementation.

**Table 1:** Measurement-domain map detailing symbols, measured parameters, and nonbulk sensor classifications.

| Symbol | Domain/access | Role in this study |
|---|---|---|
| $\chi_\Gamma$ | Open terminal Γ; $S_I = F \cdot 2e\|I\|$ proxy | Directly accessible counting object |
| $\chi_{\partial V}$ , $\chi_{div}$ | Closed ∂V ; uniform λ only | Consistency verification via Eq. (8); not an independent bulk readout |
| $\chi_{ref}$ | Model closure on ∂V = Γ ∪ Σ | Input to Eq. (12); fixed analysis convention |
| $\Delta\chi_{hidden}$ | Difference on the same Lesovik backbone | Operational definition; tested via $\Delta F_\alpha$ , $\Delta\kappa_2$ |
| $\{T_n(\varepsilon)\}$ | NEGF / edge backends | Microscopic input to Eqs. (11)–(28) |

## 2. Principles and Calculation Methods

### 2.1 Operator Gauss divergence theorem and quantum divergence theorem

This subsection extends the classical Gauss divergence theorem to the operator level and defines the generating functionals that establish the quantum divergence theorem. The dynamics of electrons is described by the Dirac field $\psi(\boldsymbol{x})$ with Lagrangian density as follows:

$$\mathcal{L} = \bar{\psi}\big(i\gamma^\mu\partial_\mu - m\big)\psi. \tag{2}$$

This Lagrangian is invariant under the global phase transformations $\psi \rightarrow e^{ie\alpha}\psi$ and $\bar{\psi} \rightarrow e^{-ie\alpha}\bar{\psi}$, where α is a constant. Noether's theorem is associated with the continuous symmetry of conserved current.

$$J^\mu = \frac{\partial\mathcal{L}}{\partial(\partial_\mu\psi)}\delta\psi + \delta\bar{\psi}\frac{\partial\mathcal{L}}{\partial(\partial_\mu\bar{\psi})} = e\bar{\psi}\gamma^\mu\psi. \tag{3}$$

After canonical quantization, the current becomes the operator $\hat{J}^\mu = e\bar{\psi}\gamma^\mu\psi$ . In the Heisenberg picture, the operator satisfies the continuity equation

$$\partial_\mu\hat{J}^\mu = 0 \Leftrightarrow \frac{\partial\hat{\rho}_e}{\partial t} + \nabla^T\hat{\boldsymbol{J}} = 0. \tag{4}$$

Integrating the continuity equation over a spatial volume V and applying the operator Gauss theorem yields

$$\frac{d}{dt}\int_V \hat{\rho}_e d^3r + \oint_{\partial V} \boldsymbol{n}^{\boldsymbol{T}} \hat{\boldsymbol{J}} dS = 0. \tag{5}$$

By introducing a classical test-function counting field $\lambda(r)$, the closed-boundary net- flux generating functional can be defined as

$$\chi_{\partial V}(\lambda) = \ln\langle \exp\left[i \oint_{\partial V} \lambda(r) \boldsymbol{n}^{\boldsymbol{T}} \hat{\boldsymbol{J}} dS\right]\rangle, \tag{6}$$

with the equivalent bulk (Gauss) form

$$\chi_{div}(\lambda) = \ln\langle \exp\left[i \int_V \lambda(r) \nabla^T \hat{\boldsymbol{J}} d^3r\right]\rangle. \tag{7}$$

Based on the operator Gauss theorem $\int_V \nabla^T \hat{\boldsymbol{J}} d^3r = \oint_{\partial V} \boldsymbol{n}^{\boldsymbol{T}} \hat{\boldsymbol{J}} dS$, the exponents coincide as operators. For a uniform $\lambda(\boldsymbol{r}) = \lambda_0$, the following identity holds:

$$\chi_{\partial V}(\lambda_0) = \chi_{\mathrm{div}}(\lambda_0). \tag{8}$$

When $\lambda(\boldsymbol{r})$ is not uniform, integration by parts yields

$$\int_V \lambda \nabla^T \hat{\boldsymbol{J}} d^3r = \oint_{\partial V} \lambda \boldsymbol{n}^{\boldsymbol{T}} \hat{\boldsymbol{J}} dS - \int_V (\nabla \lambda)^T \hat{\boldsymbol{J}} d^3r. \tag{9}$$

Therefore, $\chi_{\partial V}(\lambda) \neq \chi_{div}(\lambda)$ in general. The bulk Gauss form acquires a volume term $\propto \int_V (\nabla \lambda)^T \hat{\boldsymbol{J}} d^3r$. **Eq. (8)** expresses the uniform- $\lambda$ limit in which this term vanishes. A windowed monitoring prole $\lambda(\boldsymbol{r})$ localized on a terminal cross-section can be used to examine this split; however, this is outside the scope of the current study. Crucially, **Eq. (9)** highlights where the measurement-domain distinction introduces new information beyond the Lesovik backbone.

## 2.2 FCS, open-terminal excess, and factorization approximation

At an open cross-section $\Gamma$ of a lead, the transferred-charge operator is expressed as

$$\hat{N}_\Gamma(t) = \frac{1}{e}\int_0^1 \int_\Gamma \boldsymbol{n}^{\boldsymbol{T}} \hat{\boldsymbol{J}}(t') dS dt'. \tag{10}$$

The Levitov–Lesovik formula for transmission eigenvalues $\{T_n\}$ is

$$\chi_\Gamma(\lambda) = \sum_{n=1}^{N} \ln\left[1 + T_n\left(e^{i\lambda} - 1\right)\right]. \tag{11}$$

Expanding **Eq. (21)** to second order in $\lambda$ yields cumulants $k_1$ and $k_2$ as well as the Fano factors in **Eqs. (26)** and **(29)**.

For an open-terminal geometry, we first define a bookkeeping hidden-excess functional as

$$\Delta\chi_{hidden}(\lambda_0) \equiv \chi_\Gamma(\lambda_0) - \chi_{div}(\lambda_0), \tag{12}$$

where $\chi_{div}$ is instantiated as the closed-surface reference $\chi_{ref}$ on the same Lesovik backbone with the full monitored/hidden partition $\partial V = \Gamma \cup \Sigma$ specified. This is a bookkeeping definition: experiments access $\chi_\Gamma$ directly and $\chi_{div}$ only through the selected reference closure, not through a bulk divergence sensor, as is shown in the **Table 2.**

**Table 2**: Three full-counting-statistics generating functionals and their measurement domains.

| Object | Geometry | Definition |
|---|---|---|
| $\chi_{\partial V}(\lambda)$ | closed surface $\partial V$ | boundary flux counting |
| $\chi_{div}(\lambda)$ | bulk volume V (Gauss form) | same as $\chi_{\partial V}$ when $\lambda$ is uniform |
| $\chi_\Gamma(\lambda)$ | open section $\Gamma$ | terminal Levitov counting |

To correlate this abstract definition with observable quantities, we introduce a cumulant bridge, i.e., write $k_m^{(\alpha)} \equiv \partial_\lambda^m \ln\chi^{(\alpha)}|_{\lambda=0}$ for $\alpha \in \{\Gamma, ref\}$. Expanding **Eq. (12)** for small $\lambda$ values yields

$$\Delta\chi_{hidden}(\lambda) = \sum_{m\geq 1} \Delta k_m \frac{(i\lambda)^m}{m!}, \Delta k_m \equiv k_m^{(\Gamma)} - k_m^{(ref)}. \tag{13}$$

Terminal white-noise readout uses $F_\alpha = k_{2,\alpha}/(2|k_{1,\alpha}|)$. When gate configurations are compared at matched $|k_1|$, relative changes $\Delta F_\alpha$ track $\Delta k_2/(2|k_1|)$ to the leading order, whereas cross terminal spectra $S_{I\alpha I\beta}$ enter through mixed cumulants when $[\hat{N}_\alpha, \hat{N}_\beta] \neq 0$. **Eq. (13)** expresses the operational relationship between **Eq. (12)** and the differential protocol; it does not fix absolute $\Delta\chi_{hidden}$ values without specifying $\chi_{ref}$.

To explicitly illustrate the definitions above, we consider an analytically solvable two-channel model. On a closed surface resolving two independent channels with transmissions $T_1$ and $T_2$ at $T \ll \hbar\omega/k_B$,

$$\chi_{ref}(\lambda) = \sum_{n=1}^{2} \ln\left[1 + T_n\left(e^{i\lambda} - 1\right)\right]. \tag{14}$$

If terminal Γ monitors only channel 1, then $\chi_{\Gamma}(\lambda) = \ln[1 + T_1(e^{i\lambda} - 1)]$ and **Eq. (12)** yields

$$\Delta\chi_{hidden}(\lambda) = -\ln[1 + T_2(e^{i\lambda} - 1)]. \quad (15)$$

This implies that the hidden sector $\Sigma$ (channel 2) appears as a subtracted Lesovik factor. Under weak tunneling ($T_2 \ll 1$), **Eq. (15)** expands to $\Delta\chi_{hidden} \simeq -T_2(e^{i\lambda} - 1) + \mathcal{O}(T_2^2)$, which is consistent with the factorization scheme bound presented in **Eq. (19)**. For $C$ equal modes with $T_n = T_0$, the same factorization yields $|\mathrm{Im}\Delta\chi_{hidden}| = (C - n_{\Gamma})|\ \mathrm{Im}\ln[1 + T_0(e^{i\lambda_0} - 1)]|$ at $\lambda_0 = 0.2$.

Furthermore, when $\partial V = \Gamma \cup \Sigma$ and joint counting factorizes, either because $[\hat{J}_{\Gamma}, \hat{J}_{\Sigma}] = 0$ or the weak-tunneling limit $T_n \ll 1, \Delta\chi_{hidden}$ directly reflects the unmonitored flux on $\Sigma$. The Baker–Campbell–Hausdorff theorem presents the exact identity

$$\ln\langle e^{i\lambda_0\hat{N}_{\Gamma}} e^{i\lambda_0\hat{N}_{\Sigma}}\rangle = \ln\langle \exp\left[i\lambda_0\hat{N} + \frac{(i\lambda_0)^2}{2}[\hat{N}_{\Gamma}, \hat{N}_{\Sigma}] + \ldots\right]\rangle, \hat{N} \equiv \hat{N}_{\Gamma} + \hat{N}_{\Sigma}. \quad (16)$$

The factorization defect can be expressed as

$$\delta_{fac}(\lambda_0) \equiv \ln\langle e^{i\lambda_0\hat{N}_{\Gamma}} e^{i\lambda_0\hat{N}_{\Sigma}}\rangle - \ln\langle e^{i\lambda_0\hat{N}_{\Gamma}}\rangle - \ln\langle e^{i\lambda_0\hat{N}_{\Sigma}}\rangle. \quad (17)$$

When $[\hat{N}_{\Gamma}, \hat{N}_{\Sigma}] = 0, \delta_{fac} = 0$ identically. In the weak-tunneling limit $T_n \ll 1$, each Lesovik channel transfers one electron per window at the most, and **Eq. (11)** can be expanded as follows:

$$\ln[1 + T_n(e^{i\lambda_0} - 1)] = T_n(e^{i\lambda_0} - 1) + \mathcal{O}(T_n^2). \quad (18)$$

Thus, the cross-surface connected charge correlators are $\mathcal{O}(\overline{T}_{\Gamma}\overline{T}_{\Sigma})$, with $\overline{T}$ representing the mean transmission on each sector. Hence,

$$|\delta_{fac}(\lambda_0)| \lesssim C_{\lambda} \sum_{n,m} T_n^{(\Gamma)} T_m^{(\Sigma)}, (T_n \ll 1) \quad (19)$$

for an $O(1)$ constant $C_{\lambda}$ on the scanned $\lambda_0$ window.

## 2.3 NEGF–Lesovik transport backend

The microscopic fidelity of the Lesovik backbone expressed in **Eq. (11)** is limited by the transmission data $\{T_n(\varepsilon)\}$ supplied. When a microscopic scattering input based on the NEGF is adopted, transmission data are obtained from non-equilibrium Green's function calculations or spin-resolved edge models, whereas the bias setting, dephasing treatment, and helical edge priors remain phenomenological unless the P3 or P5 extension schemes are explicitly activated. In this mode, the preset constant $T_0$ is replaced by an energy-resolved eigenvalue spectrum computed from the NEGF or spin-resolved models. Subsequently, the same generating function and cumulant integrals for $\chi_{\partial V}$ and $\chi_{div}$ defined in **Table 2** are evaluated numerically. The complete counting chain can be summarized as

$$\mathcal{H} \overset{NEGF}{\rightarrow} \{T_n(\varepsilon)\} \overset{Eq.(21)}{\rightarrow} \chi_{\partial V} = \chi_{div} \overset{\partial_\lambda}{\rightarrow} \{k_m\} \overset{Eq.(28)}{\rightarrow} F, \tag{20}$$

with optional extension schemes P1–P5 providing microscopic $\{T_n\}$ or self-consistent bias without altering the basic formulas expressed in **Eqs. (21)–(29)**. An overview of the numerical implementation is presented in **Table 3**. Production-level benchmark calculations use P1, P2, and P4 together with the conventional dephasing treatment of **Eq. (30)** by default, unless otherwise stated.

Table 3: Microscopic extension phases P1–P5 on Levitov backbone expressed in **Eq. (20)**.

| Phase | Full name | Core function | Theoretical role |
|---|---|---|---|
| P1 | Multichannel QHE NEGF | Computes the full transmission matrix $\{T_n(\epsilon)\}$ (up to four singular values) on an IQHE strip; used for the Kumar low-temperature QPC benchmark | Validates the NEGF+Lesovik implementation in the ballistic IQHE regime |
| P2 | Helical-edge spin-resolved counting | Describes helical edges via rectangular-barrier transfer matrices with spin-resolved transmissions $T\uparrow$ and $T\downarrow$, thus introducing an effective counting weight $\zeta$ | Serves as a microscopic model for helical edges |
| P3 | Büttiker dephasing embedding | Embeds Büttiker probe self-energies | Provides a microscopic treatment of phase breaking comparable with the default |

| Phase | Full name | Core function | Theoretical role |
| --- | --- | --- | --- |
| | | $\sum_{\phi,i}^{v} = -i\eta_{\phi}/2$ directly into the NEGF retarded Green's function as an alternative to post-processing dephasing scaling (Eq. (30)) | dephasing path |
| P4 | Anderson-chain helical NEGF | Replaces the scalar disorder attenuation of P2 with spin-resolved one-dimensional tight-binding Anderson chains; matching conductance to experimental data | A more microscopic disordered model |
| P5 | Quantum Butler–Volmer coupling | Self-consistently determines the operating bias $V_{\mathrm{op}}$ via Eqs. (36)–(37) such that the Landauer conductance matches the electrochemical drive current | Bridges classical electrochemistry and quantum transport, thus closing the drive-transport loop |

Next, we reduce the operator-level description to the scattering theory. For non-interacting electrons in the steady state, the U(1) counting weight on a closed surface factorizes in the Landauer scattering basis when all lead modes crossing $\partial V$ are resolved[3, 5]. At energy $\varepsilon$, channel n contributes $\ln[1 + T_n(e^{i\lambda} - 1)]$ to the zero-temperature limit. With reservoirs at chemical potentials $\mu_L = \mu + eV_{op}/2$ and $\mu_R = \mu + eV_{op}/2$ and Fermi factors $\int_\alpha(\varepsilon) = [1 + \exp(\beta(\varepsilon - \mu_\alpha))]^{-1}$, the finite-bias, finite temperature closed-boundary generating function is

$$\chi_{\partial V}(\lambda) = \chi_{div}(\lambda) = \frac{1}{2\pi}\int d\varepsilon \sum_{n=1}^{N} \ln[1 + T_n(\varepsilon)(e^{i\lambda} - 1)f_L(1 - f_R) + T_n(\varepsilon)(e^{-i\lambda} - 1)f_R(1 - f_L)] \quad . \tag{21}$$

Open-terminal counting $\chi_\Gamma$ coincides with **Eq. (21)** only when Γ resolves the same mode set; otherwise, $\Delta\chi_{hidden} = \chi_\Gamma - \chi_{div}$ in **Eq. (12)**. In the $T \to 0$ limit with transport confined to a narrow window around $\mu$, **Eq. (21)** is reduced to **Eq. (11)**. This microscopic input mode substitutes the microscopic $T_n(\varepsilon)$ from **Eqs. (23)** and **(24)** into **Eq. (21)** and does not modify the operator identity in **Eq. (8)**. Next, we present the specific Lesovik reduction steps within the NEGF framework. Consider a noninteracting open system with a retarded Green's function as follows:

$$G^r(\varepsilon) = [(\varepsilon + i\eta)\,\mathbb{1} - \mathcal{H} - \Sigma_L - \Sigma_R]^{-1} \tag{22}$$

Here, $\mathcal{H}$ is the device Hamiltonian (modeled as a lattice IQHE strip with Peierls phases and a QPC constriction), $\Sigma_{L/R}$ denote the lead self-energies, and $\eta \to 0^+$. Broadening matrices $\Gamma_\alpha = i(\Sigma_\alpha - \Sigma_\alpha^\dagger)$ attach the left ($L$) and right ($R$) reservoirs. The transmission amplitude matrix is expressed as

$$t_{ba}(\varepsilon) = \sqrt{\Gamma_b^L} G_{ba}^r(\varepsilon) \sqrt{\Gamma_a^R}, \tag{23}$$

and the Lesovik channel transmissions are the squared singular values

$$T_n(\varepsilon) = \sigma_n^2(\varepsilon), t = U diag(\sigma_1, \ldots, \sigma_N) V^\dagger. \tag{24}$$

The Landauer conductance at chemical potential $\mu$ is expressed as $G(\mu) = (e^2/h)\Sigma_n T_n(\mu)$.

**Eq. (24)** provides the core data passed by this microscopic input mode into **Eq. (21)**; in the preset mode, $T_n(\varepsilon)$ is replaced by a constant or a Lorentzian profile associated with $T_0$.

For the lattice IQHE Hamiltonian adopted in the baseline model (Tier A), the primary NEGF device is an L × W square lattice with a nearest-neighbor hopping *t*, i.e.,

$$\mathcal{H} = \sum_{x,y} \left[-t\left(c_{x+1,y}^\dagger x_{x,y} + h.c.\right) - te^{i\phi_{x,y}}\left(c_{x,y+1}^\dagger c_{x,y} + h.c.\right) + V_{x,y} n_{x,y}\right], \tag{25}$$

Here, Peierls phases $\phi_{x,y} = B(2x+1)$ encode $B > 0$ and onsite potentials $V_{x,y}$ combine QPC constriction, wall/slit barriers, and the P1 center-plane term $V_{center}$ at the constriction midplane. Semi-infinite lead self-energies $\Sigma_{L/R}$ attach to open edge rows, thus closing **Eq. (22)**. An optional Kwant-based construction using the same geometry is available; the lattice implementation and Kwant approach are interchangeable for computing $T_n(\varepsilon)$.

Using **Eq. (21)**, we can derive the energy-resolved cumulants and Fano factor. For non-interacting channels with transmissions $\{T_n(\varepsilon)\}$, differentiating **Eq. (21)** to the second order in $\lambda$ and using $\kappa_m = \partial_\lambda^m ln\chi|_{\lambda=0}$ yields

$$\kappa_1 = \frac{1}{2\pi}\int d\varepsilon \sum_n T_n(\varepsilon)\,[f_L(\varepsilon) - f_R(\varepsilon)], \tag{26}$$

$$\kappa_2 = \frac{1}{2\pi}\int d\varepsilon \sum_n T_n\,(\varepsilon)[1 - T_n(\varepsilon)][f_L(1 - f_R) + f_R(1 - f_L)], \tag{27}$$

which are the standard Lesovik–Blanter results obtained by expanding **Eq. (21)** to the second order in $\lambda$[9,3]. The operating Fano factor is expressed as

$$F = \frac{\kappa_2}{2|\kappa_1|}, \tag{28}$$

Additionally, the white-noise shot-noise slope used throughout the mesoscopic comparisons is

$$S_I = F \cdot 2e|I|, \qquad I = e\kappa_1, \tag{29}$$

In the numerical implementation, both $\chi_{\partial V}(\lambda)$ and $\chi_{div}(\lambda)$ for uniform $\lambda$ values are both evaluated using **Eq. (21)** with the same $\{T_n(\varepsilon)\}$. This implies that **Eq. (8)** is strictly preserved at the generating- functional level under this microscopic input mode, with $\left|\chi_{\partial V} - \chi_{\mathrm{div}}\right|$ at the order of rounding errors. The cumulants agree with $\partial_\lambda^{\mathrm{m}} \ln\chi$ to machine precision.

Next, we address the treatment of dephasing and its effect on the operating point. In the default conventional path, for a length $L$ with coherence time $\tau_d$ and edge velocity $v$, $\zeta = v\tau_d$ is introduced, which suppresses each microscopic channel based on

$$T_{eff,n}(\varepsilon) = \frac{T_n(\varepsilon)}{1+(L/\zeta)^2}, \qquad L = \sqrt{A} \tag{30}$$

Here, $A$ is the electrode area. **Eq. (30)** enters **Eqs. (26)** and **(27)** after the NEGF step $T_n(\varepsilon) \to T_{eff,n}(\varepsilon)$. The bias $V_{op}$ is established by the Butler–Volmer current drive at the stated electrochemical operating point ($V_{op} \approx 0.18V, T = 298K, and\, \tau_d = 1ns$), thus yielding $F = 0.4026$ and $T_{eff} \approx 0.18$ for the dominant baseline channel, in contrast to the old preset $T_0 = 0.88$ ($F = 0.4128$, for comparison only).

Alternatively dephasing can be incorporated by embedding it directly into the retarded Green's function, instead of post-processing $T_n$, namely, the **P3** scheme. At each internal lattice site $i$, the Büttiker-probe self-energy

$$\Sigma^r_{\phi,i} = -\frac{i}{2}\eta\phi, \tag{31}$$

is added to the diagonal of **Eq. (22)**, with $\eta\phi > 0$ mapped from $\tau_d$ on the lattice time scale. Subsequently, the same $\{T_n(\varepsilon)\}$ enters **Eq. (21)** without **Eq. (30)**. The finite-bias Landauer current from an identical spectrum is expressed as

$$I = \frac{e^2}{h}\int \frac{d\varepsilon}{2\pi}\sum_n T_n(\varepsilon)\,[f_L(\varepsilon) - f_R(\varepsilon)], \tag{32}$$

with $\mu_{L/R} = \pm V_{op}/2$. P3 is implemented in the baseline lattice model. The full Keldysh lesser/greater functions and inelastic self-energies shall be investigated in future studies.

In the low-temperature multichannel case **P1**, the partition Fano factor is commonly used experimentally to describe the noise in partially open channels. When several channels are partially open, the partition Fano factor in low-temperature QPC experiments is reported to be[4-5]

$$F_{part} = \frac{\Sigma_n T_n(1-T_n)}{\Sigma_n T_n}, \tag{33}$$

which is the $T \ll \hbar\omega/k_B$ limit of **Eq. (28)** for energy-independent $\{T_n\}$. For a single dominant channel with $T_1 \ll 1$, **Eq. (33)** yields $F \to 1 - T_1$, which corresponds to Kumar noise-temperature slopes $F \to 1 - T_1$ at $T_1 \in \{3/4, 1/2, 1/4, 1/6\}$. **P1** supplies $\{T_n(\varepsilon)\}$ from the full transmission matrix expressed in **Eq. (23)** on a multirow IQHE strip. Therefore, **Eq. (33)** is evaluated with a microscopic $T_n$ instead of calibrated presets; a chiral Schrödinger fallback is invoked only when the NEGF calibration cannot reach the target $T_1$ within tolerance.

For helical edge systems, we must account for the spin-resolved counting of P2. Disordered helical edges host counter-propagating modes with spin-resolved transmissions $T_\uparrow(\varepsilon)$ and $T_\downarrow(\varepsilon)$ obtained from rectangular-barrier transfer matrices with on-site $V_\sigma$ and edge-specific disorder attenuation $\eta_\sigma \in (0,1]$. Charge monitoring on a finite spatial window introduces an effective counting weight $\zeta \in (0,1]$ (global prior, not per-sample fit). The Lesovik spectrum for terminal charge transfer is modeled as two effective channels, i.e.,

$$\left\{T_n^{(eff)}\right\} = \{\zeta T_\uparrow, \zeta T_\downarrow\}, \tag{34}$$

with conductance estimate $g = (e^2/h)(T_\uparrow + T \downarrow)$. Subsequently, **Eqs. (26)** and **(28)** define $F_{spin}$ under the same $S_I = F \cdot 2e|I|$ protocol as the chiral-QPC proxy.

To model disordered helical edges more realistically, we replace the scalar disorder attenuation with spin-resolved one-dimensional tight-binding Anderson chains, namely, the **P4** scheme. For spin $\sigma \in \{\uparrow, \downarrow\}$,

$$\mathcal{H}_\sigma = \sum_{j=1}^{N-1} \left[- t_\sigma\left(c_{j+1,\sigma}^\dagger c_{j,\sigma} + h.c.\right) + V_{j,\sigma} n_{j,\sigma}\right], \qquad (35)$$

with on-site profiles $V_{j,\sigma}$ drawn uniformly from $[-W/2, W/2]$, supplemented by a tunable rectangular CNP barrier at the chain center and independent random seeds per spin. Semi-infinite lead self-energies close a two-terminal retarded Green's function on each chain, and $T_\sigma(\varepsilon) = |t_\sigma(\varepsilon)|^2$ follows the one-dimensional (1D) reduction of **Eqs. (22)–(24)**. Conductance is matched to experimental data by grid scanning the barrier height at fixed $\varepsilon_F$.

Finally, we introduce quantum Butler–Volmer coupling to achieve a self-consistent connection between the electrochemical drive and quantum transport, **P5**. Classical electrochemistry supplies a drive current $I_{BV}(\eta)$ at overpotential $\eta = V - E_{eq}$, expressed as

$$I_{BV}(\eta) = i_0[e^{\alpha_a F_F \eta} - e^{-\alpha_a F_F \eta}] \qquad (36)$$

Here, $i_0$ is the exchange current density, $\alpha_{a,c}$ are transfer coefficients, and $F_F$ is a Faraday constant. P5 seeks $V_{op}$ such that the dimensionless Landauer conductance expressed in **Eq. (32)** matches $I_{BV}/(e^2)/h$ for the same backend $\{T_n(\varepsilon)\}$.

$$\frac{I_{BV}(V_{op} - E_{eq})}{G_0} \stackrel{!}{=} \int \frac{d\varepsilon}{2\pi} \sum_n T_n(\varepsilon)\,[f_L(\varepsilon) - f_R(\varepsilon)] \qquad (37)$$

Here, $G_0 = e^2/h$. In the implementation, the Lesovik integral $\kappa_1$ of **Eq. (26)** satisfies $I_{total} = \kappa_1(2\pi e/h)$, which is equivalent to **Eq. (32)** for the same $\{T_n(\varepsilon)\}$.

Subsequently, the converged $V_{op}$ enters **Eqs. (21)** and **(26)–(28)**, thereby closing the electrochemical drive and quantum counting in a self-consistent manner. Production-level baseline calculations use the fixed classical Butler–Volmer value $V_{op}$

≈ 0.18 V. The optional P3+P5 combined scheme, with calibrated $\eta_\phi$ ≈ 0.053 and $i_0$ anchored to the baseline bias, yields F ≈ 0.4005 at the same $V_{op}$, with $\Delta F \approx -0.0017$ relative to the default hybrid path.

For the additional complexity arising from noncommuting open cross-sections, **Sec. 2.2** discusses the bounds on the factorization defect when $[\widehat{N}_\Gamma, \widehat{N}_\Sigma] \neq 0$. The second-order Baker–Campbell–Hausdor expansion of **Eq. (16)** reiterates that $\Delta\chi_{BCH} \propto \frac{1}{2}[\widehat{N}_\Gamma, \widehat{N}_\Sigma]$ at $O(\lambda^2)$. A finite-dimensional Pauli( $||[N_\Gamma, N_\Sigma]|| = 4$ ) demonstration yields $|\Delta\chi|_{BCH} \approx 2.7 \times 10^{-3}$ at $\lambda = 0.2$; this is not a converged discretization of bosonic or chiral edge modes and must not be construed as edge-physics validation.

### 3.1 Levitov FCS, Fano factors, and open-section excess

**Figure 1** shows the implementation of the hybrid pipeline above. At the operating point ( $I \approx \frac{6.68A}{m^2}, V_{op} \approx 0.177V, T = 298\ K, and\ \tau_d = 1\ ns$ ), the Levitov benchmark integrates the energy- resolved transmission from the lattice QPC NEGF backend. After the dephasing scaling of **Eq. (30)**, a strongly sub- Poissonian Fano factor $\Delta F \approx -0.60$ is obtained relative to $F = 1$. For comparison, a legacy phenomenological preset yields another F-value, whereas an elevated- temperature model preset provides an F-value that exceeds the shot- noise limit. All values adhere to $F = \kappa_2/(2\kappa_1)$ from **Eq. (11)**.

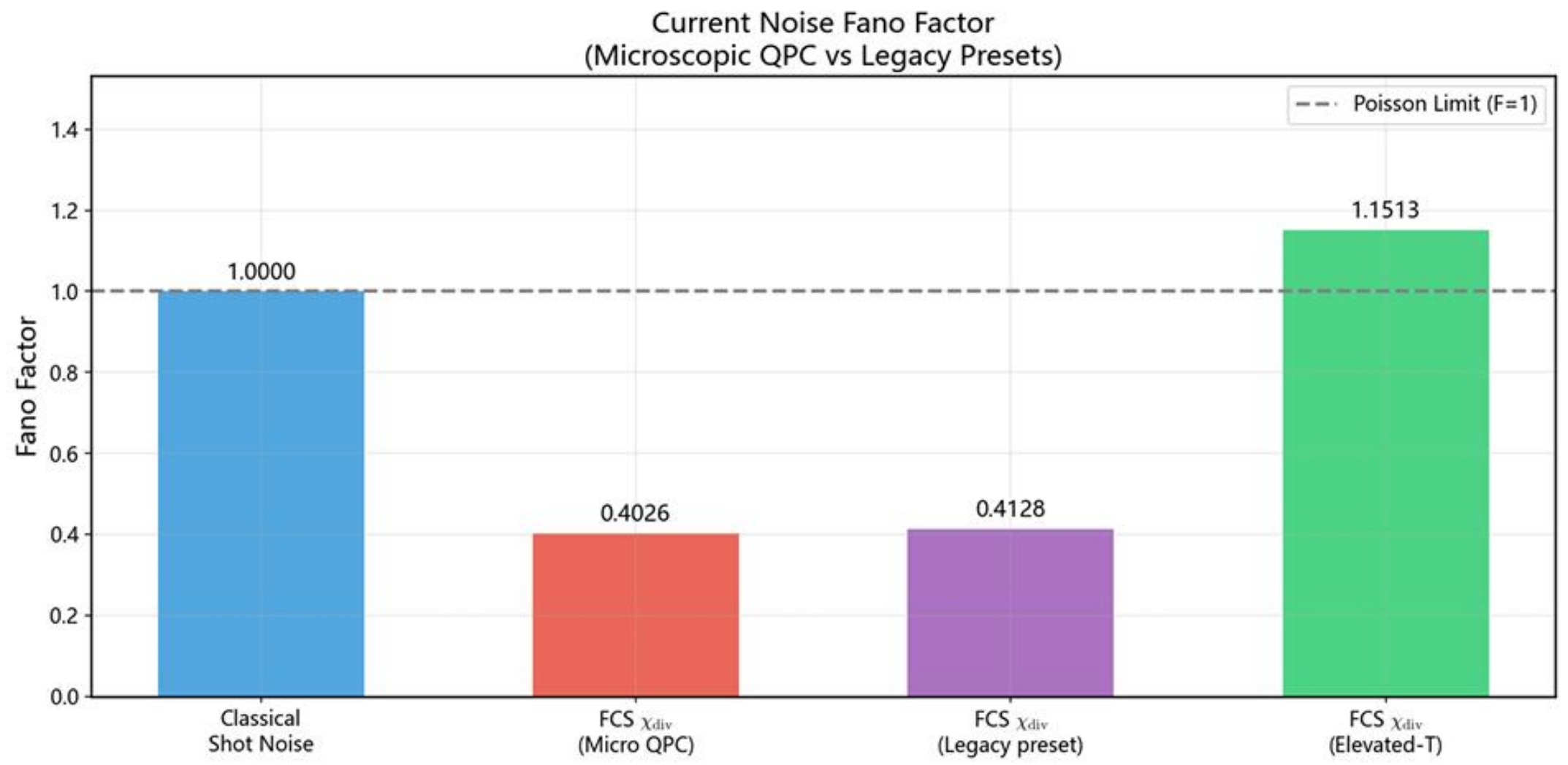

**Figure 1**: Fano factors from Levitov–Lesovik generating function. Classical shot noise ($F = 1$), 2D lattice NEGF $S(\varepsilon)$ operating point ($F = 0.4026 and\ \tau_d = 1\ ns$), legacy

preset ($F = 0.4128\ and T_0 = 0.88$), and elevated-temperature preset ($F = 1.1513$ at $400\ K$).

**Figure 2** illustrates two computational approaches for the open- section excess $\Delta\chi_{\text{hidden}} = \chi_\Gamma - \chi_{\text{div}}$ under the restrictions discussed in **Sec. 2.2**: one derived from Lesovik factorization, and the other based on the finite- difference boundary- node method.

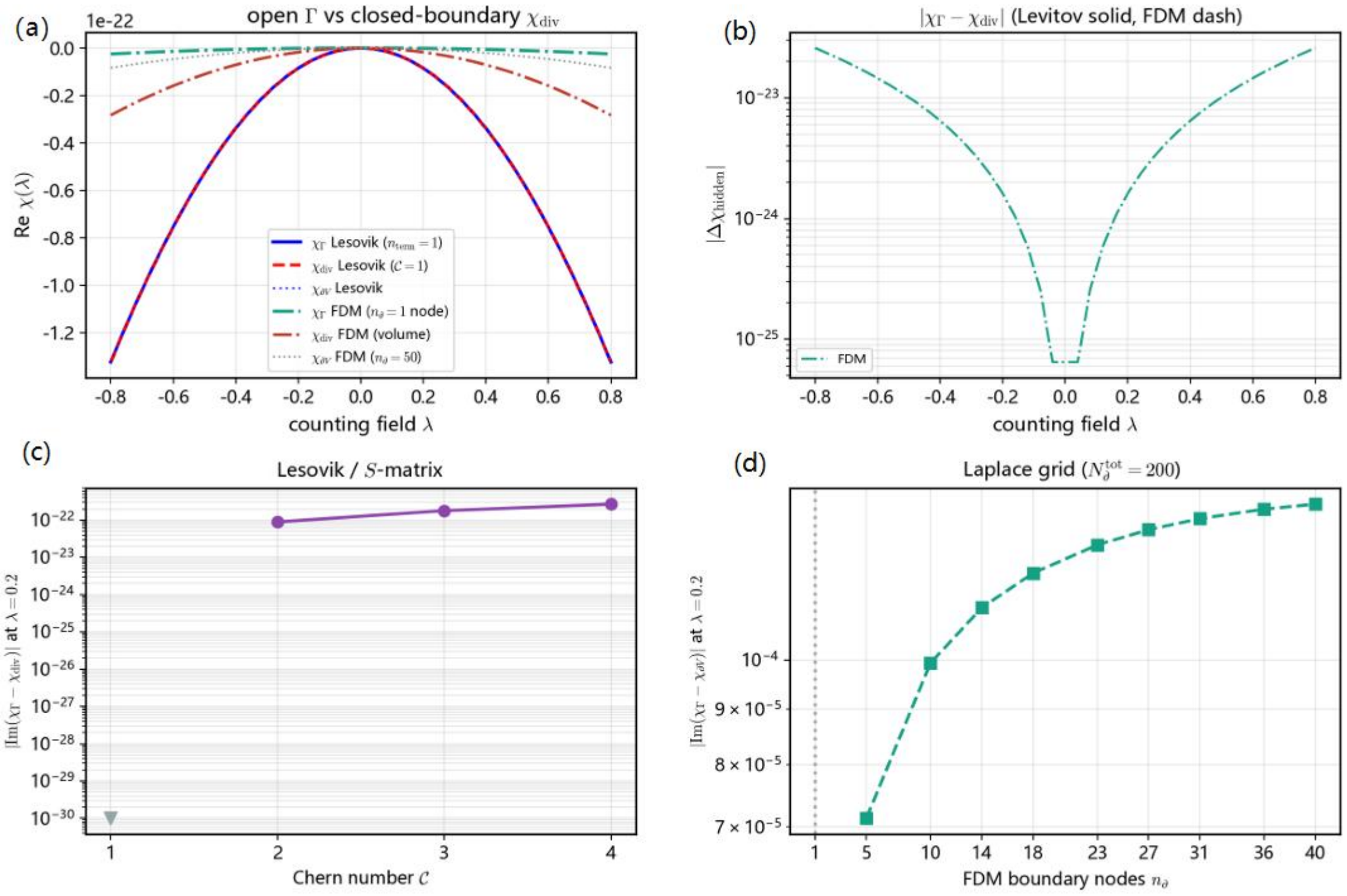

**Figure 2.** (a) Open-terminal $\chi_\Gamma(\lambda)$ vs. closed-boundary $\chi_{\text{div}}(\lambda)$ for C chiral modes (solid: Lesovik S-matrix; dash-dot: FDM boundary charges, normalized). (b) $|\Delta\chi_{\text{hidden}}| = |\chi_\Gamma - \chi_{\text{div}}|$. (c) maximum excess vs. integer Chern number C (edge mode count); $\Delta\chi \to 0$ at C = 1. (d) FDM scan vs. $n_\partial$ Laplace perimeter nodes.

### 3.2 Multiterminal geometry scan and experimental operability

**Figure 3** illustrates the change in $\Delta\chi_{hidden}(C, n_\Gamma)$ relative to the monitored- mode partition under a fixed Lesovik backbone—a direct generalization of the two- channel subtraction model. Before discussing the experimental design, the parameters measured in the laboratory should be clarified. Experiments do not measure $\chi_{div}(\lambda)$ or $\nabla \cdot \hat{j}$ in the bulk; the operationally accessible quantities are the terminal currents $I_\alpha$, low- frequency noise spectra $S_{I_\alpha I_\beta}(\omega)$, and, when available, cumulant ratios from FCS.

Under white-noise conditions, the Fano factor $F_\alpha$ obtained from $S_{I_\alpha}(0) = F_\alpha \cdot 2e|I_\alpha|$serves as the second-cumulant proxy for $\chi_\Gamma$ on terminal α.

Based on these considerations, we propose the following differential readout protocol, which does not require a direct measurement of bulk quantities. The experimental procedure comprises four steps. First, the bias and temperature are fixed on a multiterminal IQHE or a Chern-insulator device equipped with tunable splitters or QPCs. Second, for each gate configuration, the Fano factors and conductances are extracted on every monitored terminal based on the unified definition $S_I = F \cdot 2e|I|$. Third, the effective number of modes is varied either by adjusting the QPC reflectivities to split the edge channels or by changing the magnetic field; additionally, configurations with matching conductances but different monitored-mode partitions are compared—a procedure equivalent to scanning $n_\Gamma$ in **Fig. 3**. Fourth, geometry dependence is inferred from the relative changes $\Delta F_\alpha$ and cross-terminal noise correlators $S_{I_\alpha I_\beta}$between configurations, instead of by matching a single absolute Fano factor to a bulk $\chi_{div}$ value. In this protocol, gate tuning enters only through the effective transmission spectrum $\{T_n(\varepsilon)\}$ and the monitored cross-section; the reference closure $\chi_{ref}$ is fixed by the analysis code for each geometry, such that the differential $\Delta F_\alpha$ directly tracks $\Delta\kappa_2$without requiring per-sample fitting of $\Delta\chi_{hidden}$.

Theoretically, a differential protocol is intrinsically connected to a closed-boundary consistency condition. On a closed surface with uniform λ values, the identity $\chi_{\partial V} = \chi_{div}$ establishes the closed-boundary reference limit. Subsequently, open-terminal configurations compare $\chi_\Gamma$ on Γ with $\chi_{ref}$ on a selected closure $\partial V = \Gamma \cup \Sigma$. When Σ is hidden, **Eq. (15)** yields the leading Lesovik contribution to $\Delta\chi_{hidden}$, while the finite-difference boundary-node method in **Fig. 2** provides an independent lattice verification of the same excess trend.

In terms of experimental feasibility and potential confounding factors, integer quantum Hall multiterminal noise spectroscopy[4, 6] and cryogenic QPC beam splitters have already demonstrated the first two steps of the procedure above. A concrete near-term target is a GaAs/AlGaAs $\nu = 2$ Hall bar ($C = 2$) with two splitters: $n_\Gamma$ can be adjusted from 1 to 2 per terminal by the reflectivity $R$ in the $0.1-0.9$ range, thus yielding conductance steps $\Delta G \sim (e^2/h)(1-R)$. Magnetic-field sweeps access $C = 1-5$ at $B \approx 3-15T$ in standard 2DEGs. For $|I| \sim 10-100\, nA$ and $F \sim 0.3$, white-noise slopes $S_I \sim 10^{-24}-10^{-23} A^2/Hz$ require lock-in integration times $\gtrsim 1-10s$ per gate point to reach $SNR \gtrsim 3$. This is feasible but demands a differential protocol instead of absolute $\Delta\chi_{hidden}$ extraction. The confounders include Joule heating at a finite bias, contact resistance, and inelastic scattering.

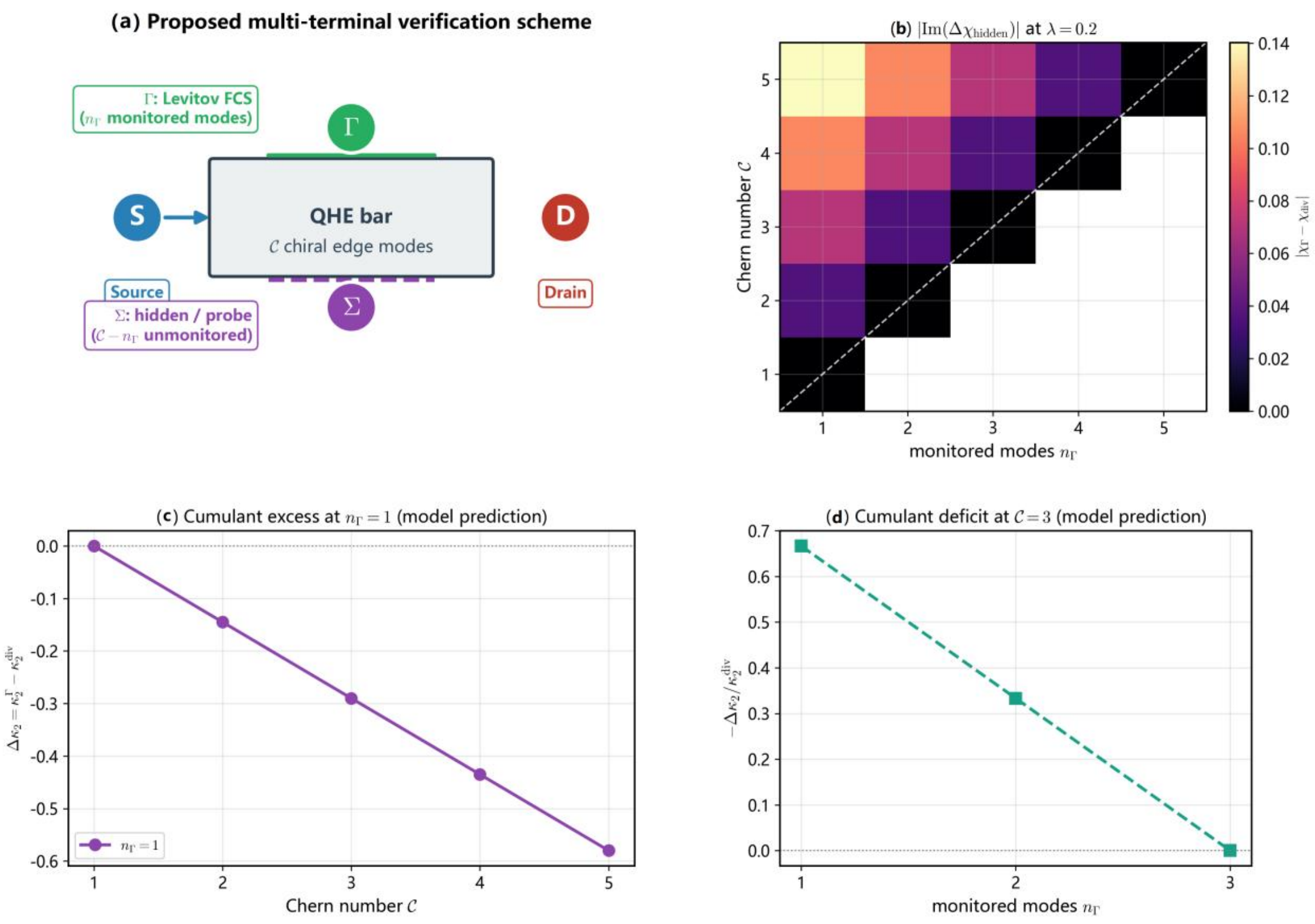

**Figure 3**: (a) Four-terminal quantum Hall bar: source S, drain D, Levitov-counted terminal Γ, and hidden auxiliary surface Σ ($C - n_\Gamma$ modes). (b) $|\mathrm{Im}(\Delta\chi_{\mathrm{hidden}})|$ at $\lambda = 0.2$ vs. Chern number C and $n_\Gamma$ from Lesovik S-matrix scan. (c) $\Delta\boldsymbol{\kappa}_2$ vs. C at $n_\Gamma = 1$. (d) Normalized cumulant deficit $-\Delta\boldsymbol{\kappa}_2/\boldsymbol{\kappa}_2^{\mathrm{div}}$ vs. $n_\Gamma$ at C = 3.

### 3.3 Mesoscopic shot-noise benchmarks: unified Levitov comparison

Regarding the open-terminal connection, low-temperature QPC and junction experiments measure only the terminal counting statistics $\chi_\Gamma(\lambda)$ on the monitored cross-section, as expressed in **Eq. (11)**. This measurement is not a single-terminal reduction of the closed-surface identity $\chi_{div} = \chi_{\partial V}$ — the open geometry breaks closure unless a reference surface $\partial V = \Gamma \cup \Sigma$ and its hidden sector are explicitly specified. Consequently, the mesoscopic data rows discussed below are intended to test the Lesovik backbone and transmission $\{T_n(\varepsilon)\}$ inputs under the unified protocol $S_I = F \cdot 2e|I|$, instead of independently validating **Eq. (12)**. Shot-noise data from the QPC, 1D wire, helical-edge, and room-temperature atomic-junction experiments were compared using a single extraction protocol and **Eq. (11)**.

First, we examined a classic low-temperature experiment on GaAs QPCs conducted by Kumar et al. In the low-temperature weak-tunneling regime[4], the partition formula $F = \sum_n T_n(1 - T_n)/\sum_n T_n$ — i.e. the Blanter–Büttiker form, which is the low-temperature limit of **Eq. (11)**—reproduces the Kumar noise-temperature slopes $F = 1 - T_1$ for channel transmissions $T_1 \in \{3/4, 1/2, 1/4, 1/6\}$ (**Fig. 3**), as well as the digitized $F(G)$ anchors in **Fig. 4** (H = 0).

The P1 multichannel NEGF calculation at $\varepsilon_F = 0.1$ —using the strengthened barrier model of **Sec. 2.3**—calibrates all four anchors of **Fig. 3**; the mesoscopic summary reports a median $|\Delta F| \approx 0.01$ on the slope rows, with no chiral QPC fallback employed (**Fig. 4**). Furthermore, the hybrid operating point of the present study ($F \approx 0.40$ at 298 K from the lattice NEGF transmission spectrum) appears between the 1/2 and 3/4 low-temperature curves (Figs. 4 and 5) on the Kumar low-temperature scale $G/(2e^2/h) \approx 1 - F \approx 0.59$; this effective transmission value is distinct from the dephased effective transmission $T_{eff} \approx 0.18$ that enters the 298 K energy integral.

Finally, regarding the alignment between theoretical calculations and literature conventions, several distinctions are noteworthy: The internal cumulants in this study use the energy- integrated definition F = $\kappa_2/(2\kappa_1)$, thus yielding $F \approx 0.40$ at 298 K with effective transmission $T_{eff} \approx 0.18$ . By contrast, the low-temperature literature comparisons employ the Blanter partition limit $F = \Sigma T_n(1 - T_n)/\Sigma T_n$ , which coincides with the noise-temperature slopes reported by Kumar et al. Both types of Fano factors originate from the same $\chi_{\text{div}}$ generating function; their correspondence connects naturally in the low-temperature limit, and a clear conversion description together with a graphical comparison is provided in **Fig. 5**.

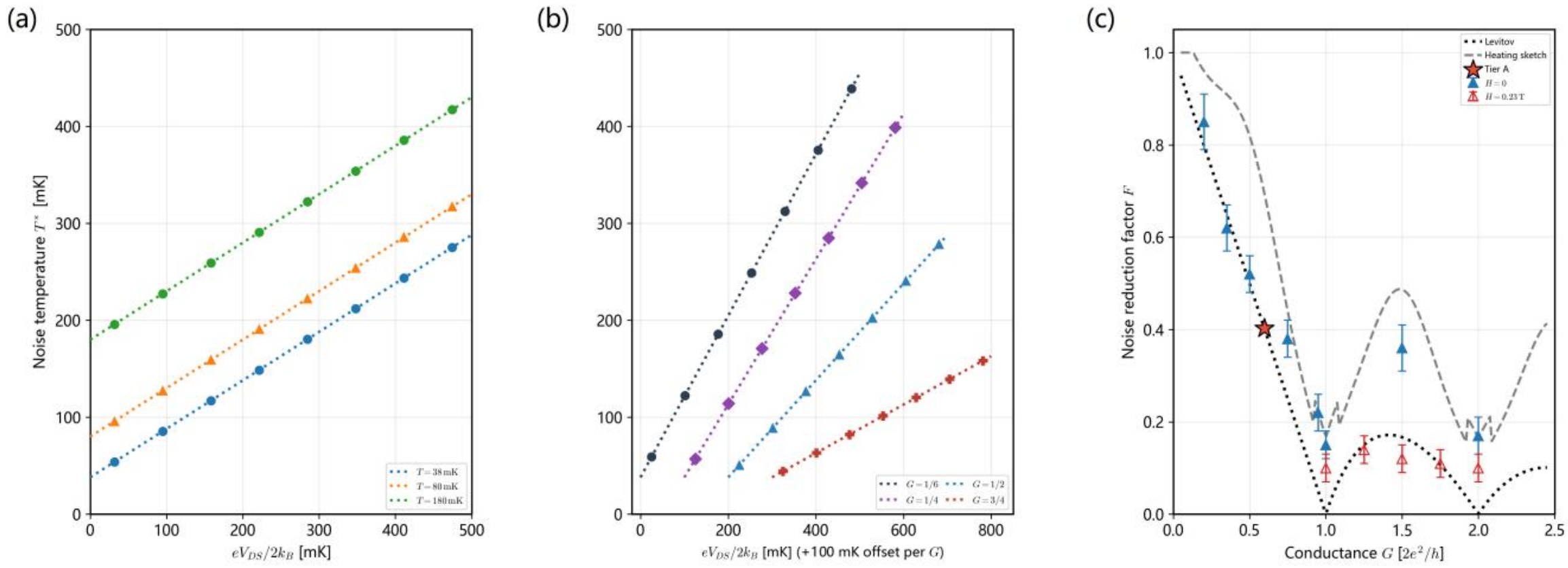

**Figure 4**: (a) Noise temperature T* vs. bias $V_{sd}$ at conductance G = 0.5 for several bath temperatures. (b) Digitized slope anchors from Fig. 3 of Kumar et al. (F = $1-T_1$), compared with P1 multichannel QHE NEGF predictions. (c) Noise-reduction factor vs. conductance, with Levitov theory, heating correction, experimental data at two magnetic fields, and Tier-A NEGF operating point.

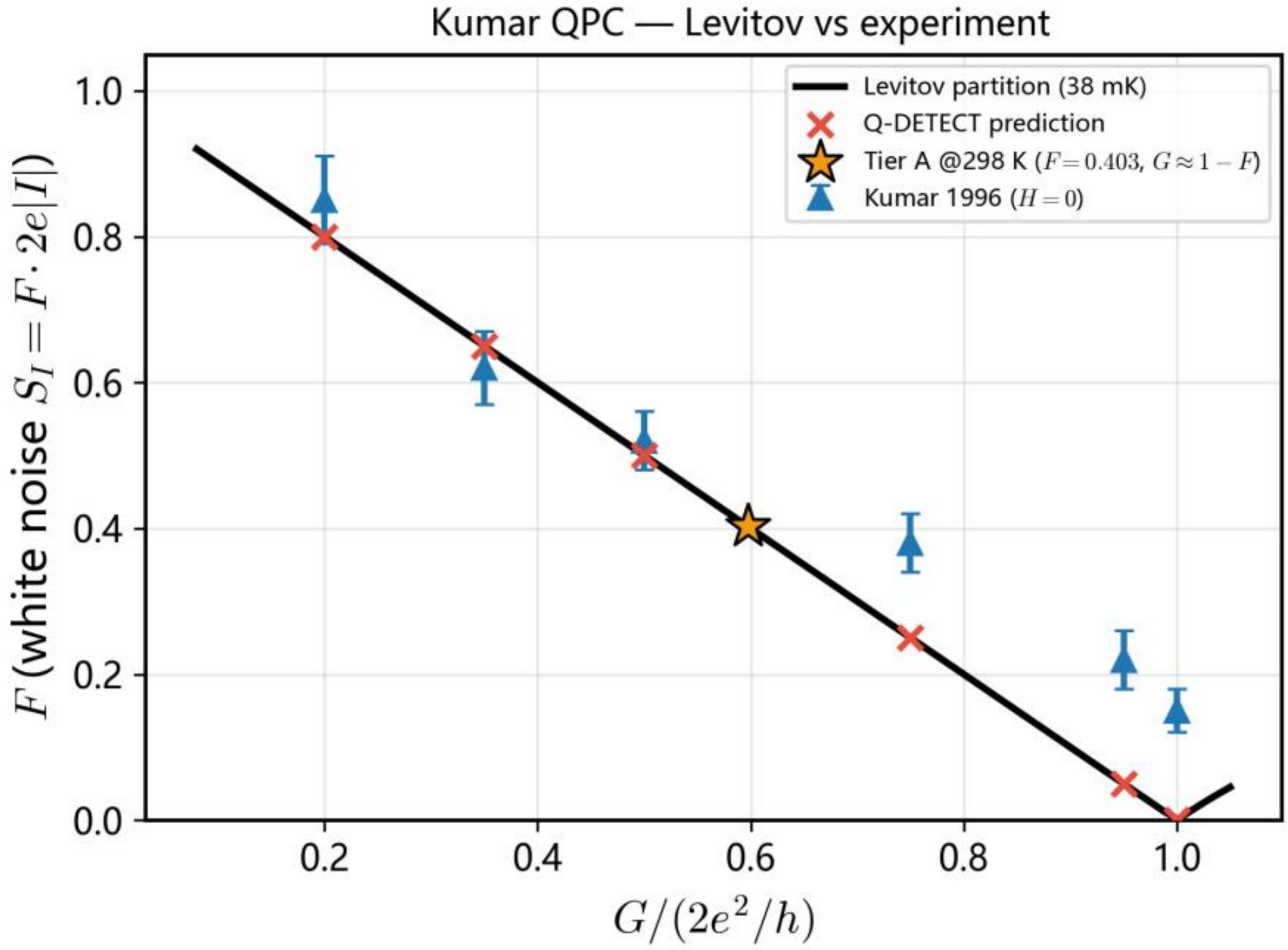

**Figure 5:** Levitov partition results obtained from P1 multichannel NEGF compared with digitized F(G) anchors; star denotes hybrid operating point on cryogenic conductance scale.

## 4 Conclusions

We established a measurement-domain framework for FCS in mesoscopic transport. The framework distinguishes three counting objects of different nature: open-terminal counting $\chi_\Gamma$, closed-boundary flux $\chi_{\partial V}$, and its Gaussian volume form $\chi_{\text{div}}$. The closed-boundary relation $\chi_{\partial V} = \chi_{\text{div}}$ is a Gauss consistency verification under a uniform counting field, not a universal open-system theorem. The principal testable proposal is a differential noise protocol that compares cumulant variations between different configurations instead of relying on absolute Fano factors, thereby partially canceling geometry-independent scattering contributions. The analytical basis of this protocol was illustrated through two-channel factorization, and a model scan of the hidden excess was provided to demonstrate its behavior across different mode partitions. Levitov counting at an open terminal measures the terminal generating function; the reference closed-surface generating function is a modeling input, not a bulk sensor. Secondary benchmarks validated numerical implementation on quantum Hall splitter systems, thus yielding sub-Poissonian Fano factors at the specified operating point and demonstrating good agreement with cryogenic QPC reference data.

**Data and Software Availability statement：**

https://github.com/Bml881020/Quantum-Divergence-Electrochemical-Topological-Diagnostics